\documentclass[aps, prd,  eqsecnum, tightenlines, notitlepage, superscriptaddress, nofootinbib, preprintnumbers, floatfix]{revtex4-1}
\pdfoutput=1
\usepackage{epsfig,amsfonts,mathrsfs,amsmath,amssymb,graphicx,color,slashed,bbm}
\usepackage{amsmath,latexsym,amssymb,graphicx,slashed,hyperref,color,enumerate,url,etoolbox}
\hypersetup{colorlinks,citecolor= nicegreen,linkcolor= nicered}
\definecolor{nicered}{rgb}{0.7,0.1,0.1}
\definecolor{nicegreen}{rgb}{0.1,0.5,0.1}

\newcommand{\lsim}{\lesssim} 
\newcommand{\gsim}{\gtrsim}  
\newcommand{\be}{\begin{equation}}
\newcommand{\ee}{\end{equation}}
\newcommand{\bea}{\begin{eqnarray}}
\newcommand{\eea}{\end{eqnarray}}
\newcommand{\MeV}{\mathrm{MeV}}
\usepackage{comment}

\def\Northwestern{Department of Physics and Astronomy,
Northwestern University, Evanston, IL 60208}
\def\Fermilab{Theoretical Physics Department, Fermilab, P.O. Box 500, Batavia, IL 60510}

\begin{document}

\title{Dark Tridents at Off-Axis Liquid Argon Neutrino Detectors
}

\author{Andr\'{e} de Gouv\^{e}a} 
\affiliation{\Northwestern}
\author{Patrick J. Fox}
\affiliation{\Fermilab}
\author{Roni Harnik}
\affiliation{\Fermilab}
\author{Kevin J. Kelly}
\affiliation{\Northwestern}
\affiliation{\Fermilab}
\author{Yue Zhang}
\affiliation{\Northwestern}
\affiliation{\Fermilab}
\date{\today}

\begin{abstract}
We present dark tridents, a new channel for exploring dark sectors in short-baseline neutrino experiments. Dark tridents are clean, distinct events where, like neutrino tridents, the scattering of a very weakly coupled particle leads to the production of a lepton--antilepton pair. Dark trident production occurs in models where long-lived dark-sector particles are produced along with the neutrinos in a beam-dump environment and interact with neutrino detectors downstream, producing an on-shell boson which decays into a pair of charged leptons.  We focus on a simple model where the dark matter particle interacts with the standard model exclusively through a dark photon, and concentrate on the region of parameter space where the dark photon mass is smaller than twice that of the dark matter particle and hence decays exclusively into standard-model particles. We compute event rates and discuss search strategies for dark tridents from dark matter at the current and upcoming liquid argon detectors aligned with the Booster beam at Fermilab -- MicroBooNE, SBND, and ICARUS -- assuming the dark sector particles are produced off-axis in the higher energy NuMI beam. We find that MicroBooNE has already recorded enough data to be competitive with existing bounds on this dark sector model, and that new regions of parameter space will be probed with future data and experiments. 
\end{abstract}

\preprint{FERMILAB-PUB-18-433-T, NUHEP-TH/18-10}

\maketitle

\setcounter{equation}{0}
\section{Introduction}
\label{sec:introduction}

The nature of dark matter remains a fascinating puzzle for particle physics and cosmology. Its resolution seems to point, almost unambiguously, to the existence of degrees of freedom not accounted for in the standard model of particle physics. On the particle physics side, terrestrial and astrophysical searches for a dark matter fundamental particle have yet to bear any fruit. The current situation strongly suggests broadening the search strategy, exploring options outside of the traditional ones. On a separate front, the discovery of neutrino masses has also demonstrated that the standard model is incomplete. Many experiments have been, and are being, built in order to fully explore the new physics revealed by neutrino oscillation experiments.  Over the past several years, experiments that make use of intense neutrino beams have been exploited as potential laboratories for dark matter searches ~\cite{Batell:2009di, Essig:2010gu, deNiverville:2011it, Dobrescu:2014ita, Coloma:2015pih, Aguilar-Arevalo:2017mqx, Frugiuele:2017zvx}. Most of these proposals and searches focused on neutral-current-like dark matter--nucleus scattering in the near detectors of present and future neutrino experiments. The current and next generation of neutrino detectors, particularly those using liquid argon technology, boast excellent tracking, particle ID, and resolution, and are capable of conducting a broader set of searches for new phenomena (see~\cite{Izaguirre:2017bqb, Magill:2018jla, Magill:2018tbb, deNiverville:2018dbu} for some recent proposals). This paper aims to further explore these capabilities.

\begin{figure}
\centering
\includegraphics[width=0.7\textwidth]{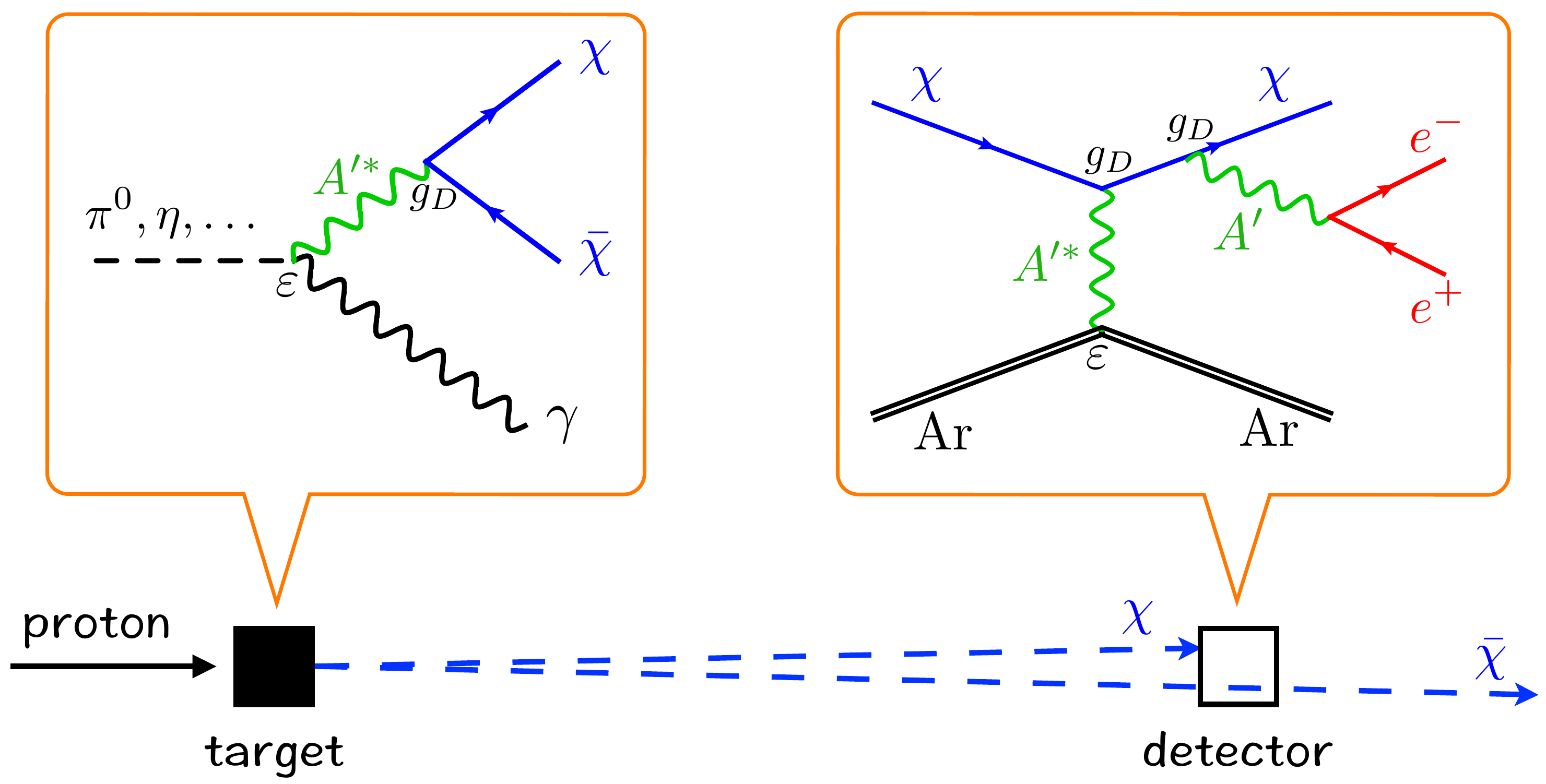}
\caption{Dark trident production in the benchmark dark sector model, described in the text. Protons on target produce mesons whose decays produce pairs of dark matter particles, $\chi$ and $\bar{\chi}$. Some of the dark matter particles will reach the detector (not to scale) and will scatter with the argon nuclei that fill the detector. The scattering event can lead to the emission of one (or several) dark photon(s) $A^\prime$, which decays to a pair of charged leptons (electrons in the case of the figure).}
\label{fig:proddet}
\end{figure}

We introduce and discuss the \emph{dark trident}, a new channel for searches for light and weakly-coupled dark sector particles in neutrino-beam experiments.  The signal, depicted in Figure~\ref{fig:proddet}, consists of the production of one or more lepton--antilepton resonances in a near detector. This is not unlike neutrino trident production, $\nu+N\to \nu+N+\ell^++\ell^-$, where $N$ is some nuclear state and $\ell$ is a charged lepton, hence the denomination. The dark trident topology is expected to be clean and, under the right circumstances, the irreducible standard model backgrounds are very small. We discuss these issues in detail in Section~\ref{sec:signal}. We comment on multi-trident production in Section~\ref{sec:multi}.

Searching for new physics in dilepton resonances in fixed target experiments is commonplace~\cite{Bjorken:2009mm, Abrahamyan:2011gv, Battaglieri:2014hga, Berlin:2018pwi}. In the simplest models, the new resonance is produced at the target and decays inside or near an instrumented detector volume. Neutrino facilities, where the near detector is usually placed hundreds of meters downstream of the target, have sensitivity to regions of parameter space where the new resonance is long-lived enough to reach the detector. This implies small couplings and in turn suppressed production rates (though some sensitivity remains if the backgrounds are low~\cite{Essig:2010gu}). Probing models with stronger couplings, where the typical decay lengths are of order a few meters, requires detectors located very close to the target region (see for example~\cite{Berlin:2018pwi}) while probing models where the decay is even more prompt requires different experimental strategies, such as thin targets (see, for example,~\cite{Bjorken:2009mm, Abrahamyan:2011gv, Battaglieri:2014hga}). 

Here, instead, we explore a different scenario, motivated by the possibility that the dark matter particle (or one of the dark matter particles) is a light weakly-interacting massive particle (we consider masses between an MeV and a GeV). In this case, if the dark matter particle is a thermal relic -- i.e., if at any point in the universe's history it was in thermal equilibrium with the standard model particles --  one must also postulate the existence of new interactions associated to equally light mediators (see e.g.~\cite{Pospelov:2007mp, Feng:2008ya}).\footnote{Dark matter models with light mediators have also been proposed in other contexts~\cite{Pospelov:2008jd, ArkaniHamed:2008qn, Rocha:2012jg}.} We consider the simplest such model, described in detail in Section~\ref{sec:amodel}. The dark matter particle is a standard model singlet charged under a new $U(1)_D$ gauge interaction. None of the standard model particles are charged under $U(1)_D$ but, assuming the  $U(1)_D$ symmetry is broken, kinetic mixing allows the new gauge boson -- the dark photon -- to interact with charged standard model particles. 

In this scenario,\footnote{This type of signal is not endemic to dark matter models. Frameworks in which a heavy sterile neutrino is produced in neutrino beams and then proceeds to decay to a dark photon have been recently proposed to explain the MiniBooNE low energy excess \cite{Bertuzzo:2018itn,Ballett:2018ynz}.  These also lead to similar phenomenology.} it is possible to produce dark photons in the near detectors of neutrino beam experiments and have the dark photon decay predominantly and promptly into a lepton--antilepton pair, as depicted in Figure~\ref{fig:proddet}.  First, a dark matter particle $\chi$ is produced at the target. It is very long-lived and weakly interacting so it easily finds its way to the detector unperturbed, regardless of how far away it is. Second, the dark photon $A'$ is produced on-shell by in the collision of a dark matter particle with a nucleus. The dark photon decays promptly in the detector. The relevance of the dark trident channel depends dramatically on the parameters of the model. First, the branching ratio of the dark photon into leptons must be large. This can only be achieved if the decay of the dark photon into dark sector particles, including the dark matter particle, is forbidden, i.e., $m_{A'}<2m_{\chi}$. Second, the probability of emitting a dark photon during $\chi$ scattering must be large. This requires the dark coupling constant to be large or the dark photon mass to be very small. Third, the dark matter production rate should not be too small. Here, the condition $m_{A'}<2m_{\chi}$ is a hindrance as it implies that dark matter production involves off-shell dark photon intermediate states. The literature reveals~\cite{Batell:2014mga,Kahn:2014sra} that it is possible to produce enough dark matter particles when the dark photons are off-shell, an issue we discuss in detail in Sec.~\ref{sec:beam}. Previous studies considered only leading-order dark-matter--nucleus scattering as the detection signal. One narrow interpretation of our contribution is that we are exploring the benefits of the rarer, but much more distinct, detection signal in the region of parameter space where dark matter production is suppressed.   

The dark trident signal provides ample experimental handles. In the remainder of this manuscript, we study the phenomenology of the signal and show that, for our benchmark dark matter model, off-axis liquid argon detectors, including the ones currently running or under construction at Fermilab,  have improved sensitivity because neutrino backgrounds can be suppressed~\cite{Coloma:2015pih}.  The kinematics of dark trident events, including angular distributions, invariant mass, as well as particle identification will be useful in further suppressing backgrounds.

\section{A Benchmark Model}
\label{sec:amodel}

We consider the model where a fermionic dark matter particle $\chi$ with mass $M_{\chi}$ interacts with those in the standard model through a massive dark photon $A'$ with mass $M_{A'}$ and its kinetic mixing with the photon. The Lagrangian takes the form 
\begin{eqnarray}\label{Lvector}
\mathcal{L} = \mathcal{L}_{\rm SM} + \bar\chi i\gamma^\mu(\partial_\mu - i g_D A'_\mu) \chi - M_\chi \bar \chi \chi  
 -\frac{1}{4} F'_{\mu\nu} F'^{\mu\nu} - \frac{\varepsilon}{2} F_{\mu\nu} F'^{\mu\nu} + \frac{1}{2} M^2_{A'} A'_\mu A'^\mu \ ,
\end{eqnarray}
where $\varepsilon$ is the kinetic mixing parameter and $g_D$ is the gauge coupling associated to the dark $U(1)_D$. We define the dark fine-structure constant $\alpha_D\equiv g_D^2/(4\pi)$. The new interactions are such that $\chi$ can only be pair produced, rendering it absolutely stable and a dark matter candidate. 

\begin{figure}
\centering
\includegraphics[width=0.8\linewidth]{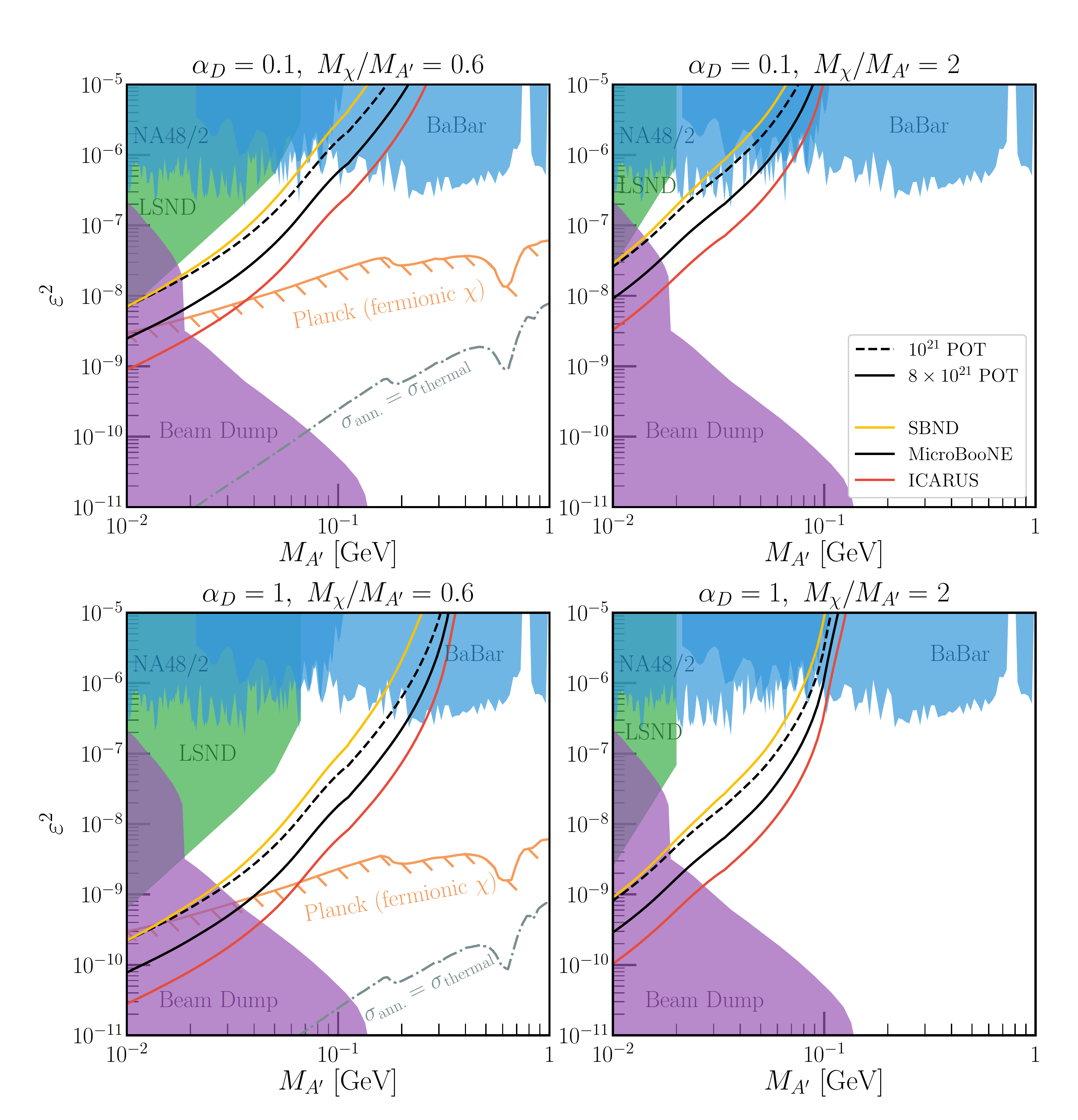}
\caption{Solid lines: regions of $M_{A^\prime}$ vs. $\varepsilon^2$ parameter space for which we expect $10$ signal events after $8\times 10^{21}$ protons on the NuMI target have been accumulated. The gold line corresponds to the SBND detector, the black line to the MicroBooNE detector, and the red line to the ICARUS detector. The dashed black line corresponds to the region for which $10$ signal events are expected at MicroBooNE, assuming $10^{21}$ POT. The top (bottom) row corresponds to $\alpha_D = 0.1$ ($\alpha_D = 1$). The left (right) panels corresponds to $M_\chi = 0.6\,M_{A^\prime}$ ($M_\chi = 2 M_{A^\prime}$). Here we assume the dark matter relic abundance is symmetric. See text for more detail regarding existing limits.}
\label{fig:MicroBooNERegions}
\end{figure}

\begin{figure}
\centering
\includegraphics[width=0.8\linewidth]{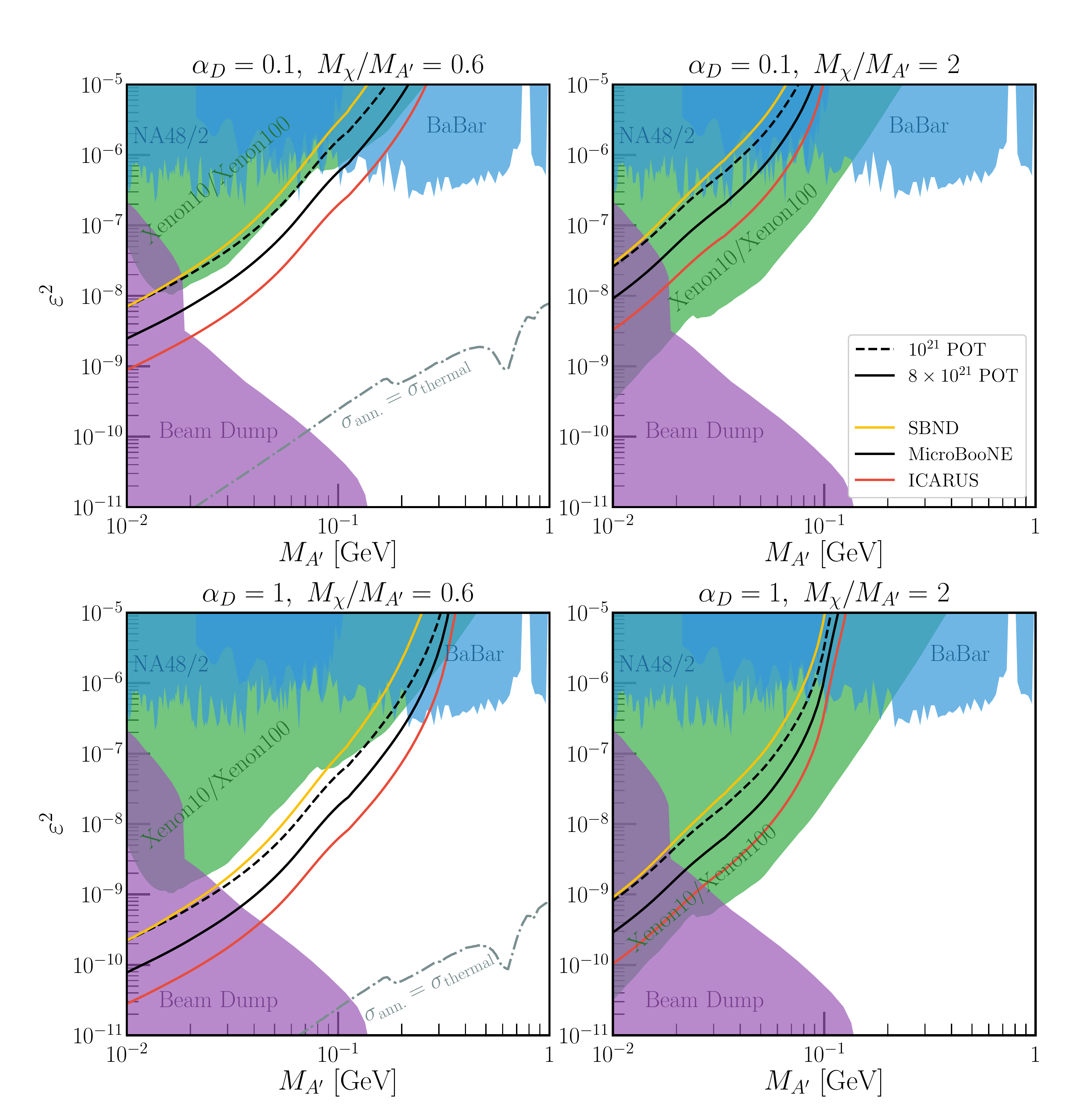}
\caption{Similar to Fig.~\ref{fig:MicroBooNERegions} but for the case of asymmetric dark matter. 
In order to efficiently deplete the symmetric relic density, the annihilation cross section needs to be above the thermal one. 
For the left panels, this corresponds to values of $\varepsilon$ larger than those of the thermal (blue, dot-dashed) curves.
The xenon direct detection experiments excluded the green-colored regions.}
\label{fig:MicroBooNERegionsADM}
\end{figure}

As far as the relic density of the $\chi$ particle is concerned, when $M_{A'}> M_\chi$, $\chi$ annihilates exclusively to SM fermions through the kinetically mixed dark photon. The relic density is set by the dark matter annihilation cross section,
\begin{equation}
\sigma_{\rm ann} v = \frac{6 g_D^2 M_\chi}{(4M_\chi^2 - M_{A'}^2)^2} \Gamma_{A'}(2M_{\chi}) \ ,
\end{equation}
where $\Gamma_{A'}(2M_{\chi})$ is the decay width of $A'$ to SM fermions, evaluated when its mass is equal to $2M_\chi$, which in turn is related to the $R$ ratio. Figs.~\ref{fig:MicroBooNERegions} (left) and \ref{fig:MicroBooNERegionsADM} (left) depict -- gray dot-dashed line -- the region of $M_{A'}\times \varepsilon^2$ parameter space where the thermal relic abundance of $\chi$ makes up all of the dark matter, for $M_{\chi}=0.6M_{A'}$ and $\alpha_D=0.1$ (top) or $1$ (bottom). For points in the parameters space that lie above these lines, the thermal relic density of $\chi$ particles is too small to make up all of the dark matter. Nonetheless, because we are interested in small (sub-GeV) dark matter masses, there are strong constraints on $\chi+\bar{\chi}$--annihilations during the formation of the cosmic microwave background (CMB)~\cite{Slatyer:2015jla}. The Planck experiment sets a lower limit on $\varepsilon$ which excludes the orange-line-bounded region in Fig.~\ref{fig:MicroBooNERegions} (left)~\cite{Slatyer:2009yq,Madhavacheril:2013cna}. These CMB constraints, however, would be absent if the dark matter candidate were a boson instead of a fermion, since the annihilation is $p$-wave suppressed in the case of a boson.  

When $M_{A'}<M_\chi$, $\chi$'s can annihilate to two dark photons and, in the limit where $\varepsilon$ is small, the relic density is exclusively determined by the parameters of the dark sector. This scenario is known as secluded dark matter~\cite{Pospelov:2007mp}. Quantitatively, in the non-relativistic limit 
\begin{equation}
\sigma_{\rm ann} v = \frac{g_D^4}{16\pi M_\chi^2}\frac{\left(1-\frac{M_{A'}^2}{M_\chi^2}\right)^{3/2}}{\left(1-\frac{M_{A'}^2}{2M_\chi^2}\right)^2}~.
\end{equation}
For the annihilation cross section at arbitrary center-of-mass energy, see \cite{Cline:2014dwa}.
For the dark-sector parameters of interest here -- large $g_D$, sub-GeV masses --  this annihilation cross-section is very large and the thermal relic density of $\chi$ particles is always very small. For this reason, there is no dot-dashed line or orange line in the right-hand panels of Figs.~\ref{fig:MicroBooNERegions} and \ref{fig:MicroBooNERegionsADM}, which depict the region of $M_{A'}\times \varepsilon^2$ parameter space where $M_{\chi}=2M_{A'}$.\footnote{As pointed out in \cite{DAgnolo:2015ujb, Cline:2017tka}, for very light dark-sector particles, this annihilation channel may efficiently deplete the dark matter relic abundance even when $M_{\chi}<M_{A'}$ and $\chi\bar{\chi}\rightarrow A' A'$ annihilation relies on non-zero temperature effects. This is significant when the mass ratio $M_{\chi}/M_{A'}$ is close enough to one and the dark photon quickly and dominantly decays into standard model particles, which can be satisfied even for very small $\varepsilon^2 \sim 10^{-14}$. For the mass ratios $M_{\chi}/M_{A'}$ chosen in the left-hand panels of Figs.~\ref{fig:MicroBooNERegions} and \ref{fig:MicroBooNERegionsADM}, these concerns do not apply.}

We are interested in the region of parameter space where the thermal relic abundance of $\chi$ is just right or relatively too small. In this case, $\chi$ either makes up only a fraction of the dark matter of the universe or the $\chi$ relic abundance is not trivially related to its thermal relic abundance. For example, it could be a consequence of a dynamically generated $\chi$--$\bar{\chi}$ asymmetry: asymmetric dark matter~\cite{Nussinov:1985xr,Kaplan:2009ag}. In this case, the ``larger-than-thermal'' cross section may be important for removing the symmetric dark matter component. If $\chi$ does constitute all of the dark matter, direct detection limits with electron scattering from XENON10~\cite{Essig:2012yx} and XENON100~\cite{Essig:2017kqs} apply. These are depicted as green-shaded regions in Fig.~\ref{fig:MicroBooNERegionsADM}. The distinction between Figs.~\ref{fig:MicroBooNERegions} and \ref{fig:MicroBooNERegionsADM} is as follows. In Fig.~\ref{fig:MicroBooNERegions}, we assume $\chi$ to be a thermal, ``symmetric'' relic and, for the parameters of interest here, $\chi$ makes up only a small fraction of the dark matter. In Fig.~\ref{fig:MicroBooNERegionsADM}, we assume $\chi$ to be asymmetric dark matter and assuming it makes up all the dark matter, direct detection bounds are formidable competitors to the signal we will ultimately discuss here. 

As far as laboratory constraints are concerned, it is important to distinguish whether the dark photon can or cannot decay to dark matter. When $M_{A'}> 2M_\chi$, dark matter can be produced by $A'$ decays. This can lead to relatively intense dark matter beams in fixed target setups~\cite{Batell:2009di}, allowing for interesting dark matter search strategies~\cite{deNiverville:2011it, Dobrescu:2014ita, Coloma:2015pih, Aguilar-Arevalo:2017mqx, Frugiuele:2017zvx}. If, instead, $M_{A'}<2M_\chi$, dark matter is, phenomenologically speaking, darker since its production requires an off-shell dark photon. This region is the focus of this work. There are several existing laboratory constraints on this model, including limits from rare pion decays at NA48/2~\cite{Batley:2015lha}, beam dump experiments such as E137~\cite{Bjorken:1988as}, E141~\cite{Riordan:1987aw}, and E774~\cite{Bross:1989mp}, as well as searches for promptly decaying dark photon events at BaBar~\cite{Lees:2014xha}. These translate into the labelled purple and blue excluded regions of the $M_{A'}\times \varepsilon^2$ parameter space, depicted in Figs.~\ref{fig:MicroBooNERegions} and \ref{fig:MicroBooNERegionsADM}.

Previous studies have considered DM production through an off-shell $A'$ at LSND~\cite{Kahn:2014sra} and E137 at SLAC~\cite{Batell:2014mga}, with detection occurring through a neutral-current-like scattering.  As the beam energy at LSND was only 800 MeV, DM production occurs overwhelmingly through $\pi^0$ decay and the bounds are weak for higher $A'$ and $\chi$ masses \cite{Kahn:2014sra}.  The beam energy at E137 \cite{Batell:2014mga} was 20 GeV and DM pairs were produced mostly through bremsstrahlung.  The E137 bounds are subdominant to those coming from LSND. The bounds from~\cite{Kahn:2014sra} are depicted as green-shaded regions in Fig.~\ref{fig:MicroBooNERegions}\footnote{In comparison with Ref.~\cite{Kahn:2014sra}, we include a scaling factor on the bounds from LSND (on the parameter $\varepsilon^2$) of 3.62 for $\alpha_D = 0.1$ to correct for errors in previous works~\cite{GordanPrivateCommunication}. This conversion factor is determined by comparing Refs.~\cite{Kahn:2014sra} and~\cite{deNiverville:2018dbu}.}. 
We omit these curves from Fig.~\ref{fig:MicroBooNERegionsADM}, since they are surpassed by those from direct detection.

In Figs.~\ref{fig:MicroBooNERegions} and \ref{fig:MicroBooNERegionsADM} we present the outcome of our dark matter trident analysis, the details of which are contained in the following sections.  In each case, the region above the solid black, gold, and red line corresponds to the part of parameter space where we expect MicroBooNE, SBND, and ICARUS, respectively, to produce 10 or more dark trident events after $8\times 10^{21}$ protons on target (POT) delivered.  We assume in these plots that the detector efficiency is 100\%.  We address this assumption below.  In the case of MicroBooNE, we also show the region with at least 10 events after $10^{21}$ POT, which is the amount of data the experiment has already recorded from the NuMI beam.  We see that, with the exception of a non-thermal dark matter relic with $M_\chi/M_{A'}=2$, an analysis of existing MicroBooNE data will  supersede the present experimental constraints and in some cases future analysis can even close the gap to the cosmological constraints, ruling out completely a range of dark matter masses.

\section{Off-Axis Dark Matter Beam}
\label{sec:beam}

In a proton fixed-target environment, light dark matter (masses below several hundred MeV) is most efficiently produced in the decay of neutral hadrons. These facilities benefit from a large number of high energy POT, each of which produces several neutral mesons. For the parameter space of interest here ($M_{A'}<2M_\chi$, $M_{A'}\lesssim 100$~MeV), the most important production channels involve light neutral mesons, 
see the Feynman diagram in Fig.~\ref{fig:proddet} (left),
\begin{equation}\label{eq:MainProduction}
\pi^0 \to \gamma \chi \bar\chi, \ \ \ \ \ \eta \to \gamma \chi \bar\chi \ .
\end{equation}
These processes occur via an off-shell dark photon $A'$ and its kinetic mixing with the photon. The production via $\rho$-meson decay, $\rho^0\to\chi\bar\chi$, is less important given its very broad width. In addition to meson decay, dark matter particles are also pair-produced via the Drell-Yan process. For center-of-mass energies above $\sim 1.7$\,GeV, this production rate can be safely calculated using the parton model. Given the masses and couplings of interest here, we find the Drell-Yan production rate to be several orders of magnitude smaller than that of meson decays and will not include the contribution of this channel.

The dark matter beam produced in our scenario is significantly broader than the neutrino beam. This is due to two effects. (1) The parent particles of the neutrinos are charged pions/kaons, which are focused by the magnetic horns.  More precisely, the magnetic horns only focus one charge of pions while the other sign is defocussed; the net effect is still to increase the forward neutrino flux.  The parents of the $\chi$ particles are neutral mesons which are unaffected by the magnetic horn.  (2) Neutrinos are predominantly produced in two-body decays (e.g. $\pi^+ \to \mu^++\nu$), while the $\chi$'s are produced in three-body decays. As a result, detectors located away from the proton beam axis will have a larger signal to background ratio and potentially improved reach. This preference for off-axis detectors is a useful approach to reducing the beam-induced neutrino backgrounds in a large class of exotic searches.

\begin{figure}
\centering
\includegraphics[width=0.8\textwidth]{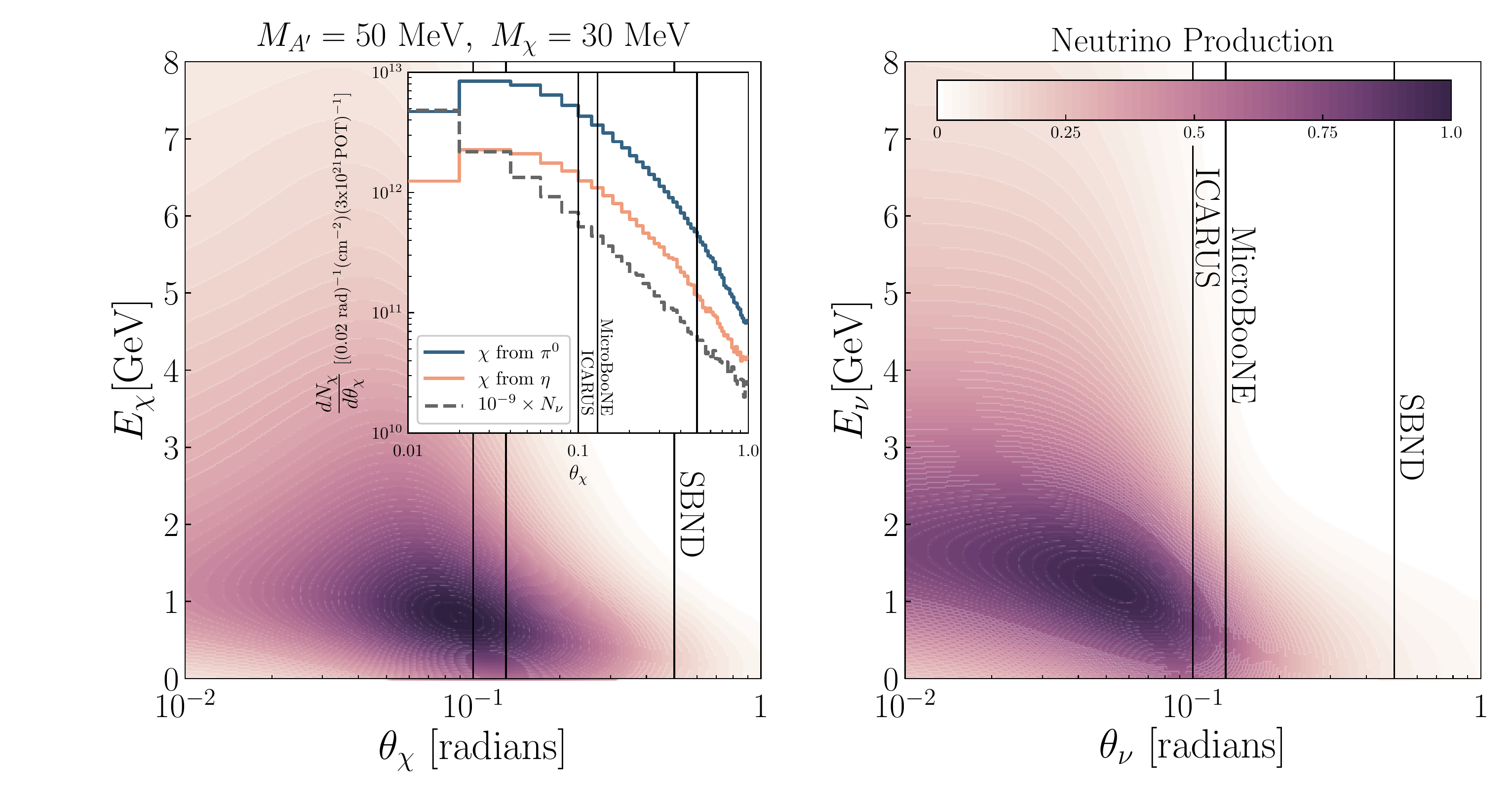}
\caption{The differential distribution of $\chi$ particle (left) and neutrino (right) production as a function of the particle energy and production angle in arbitrary units. The model parameters used here are $\varepsilon=10^{-3}$, $M_\chi=30\,$MeV and $M_{A'}=50$\,MeV.
In the insert we show the energy-integrated $\chi$ particle flux as a function of the production angle, 684\,meters away from the NuMI target, due to $\pi^0$ (blue) and $\eta$ (orange) decays as well as the off-axis angular distribution of the neutrino flux (gray).  The neutrino flux has been rescaled by $10^{-9}$.
A larger fraction of $\chi$'s are produced off-axis as compared to neutrinos. 
}
\label{fig:ETheta}
\end{figure}

Quantitatively, we consider the NuMI beam at Fermilab -- a $120$ GeV proton beam striking a graphite target -- and simulate the production of meson $\mathfrak{m}=\pi^0, \eta$ using {\tt PYTHIA8}. For every POT, roughly $c_{\pi^0}=4.5$ $\pi^0$s and $c_\eta=0.5$ $\eta$ mesons are produced.  As a result,
the number of dark matter particles produced for each parent meson channel is
\be
N_{\chi} \simeq 2 c_{\mathfrak{m}} Br(\mathfrak{m}\rightarrow \gamma\gamma) \varepsilon^2\alpha_D N_\mathrm{POT} I\left(\frac{M^2_{\chi}}{m^2_{\mathfrak{m}}}, \frac{M_{A'}^2}{m^2_{\mathfrak{m}}}\right)\, ,	
\ee
where the dimensionless function $I(x,y)$ is related to the ratio of partial widths $\Gamma(\mathfrak{m}\rightarrow \gamma\bar\chi\chi)/\Gamma(\mathfrak{m}\rightarrow \gamma\gamma)$. Explicitly,  
\bea
I(x,y) = \frac{2}{3\pi}\int_{4x}^1 dz \sqrt{1-\frac{4x}{z}}\frac{1-z}{(z-y)^2}
\left(12x^3+6x^2(3z-2) + x(5z-2)(z-1)+z(z-1)^2\right)\,.
\eea
This function is $\mathcal{O}(1)$ over most of the parameter range of interest. In the limit $m_{\mathfrak{m}} \gg 2 M_{\chi} \gtrsim M_{A'}$, however, $I\sim \log m_\mathfrak{m}^2/M_\chi^2$ due to soft or collinear infrared effects.
%
%
We simulate the NuMI dark matter angular distributions and energy spectra from  $\pi^0, \eta$ decays on an event-by-event basis. The double-differential distributions are depicted in Fig.~\ref{fig:ETheta} (left), for $\varepsilon=10^{-3}$, $M_\chi=30\,$MeV and $M_{A'}=50$\,MeV. 
Here $\theta_{\chi}$ is the off-axis angle and $E_\chi$ is the dark matter energy, both in the lab frame. The $\chi$ angular distribution is largely insensitive to $M_\chi$ as long as it is well below the kinematic threshold of the decay. 

For comparison, we produced a rough estimate of the angle and energy distribution for the neutrino beam, depicted in Fig.~\ref{fig:ETheta} (right). Neutrinos are produced mainly from the decay of charged pions and kaons, one sign of which are focused in the forward direction.  We estimate the double-differential distribution of the neutrino flux as follows: we obtain the positively charged pion and kaon energies and momenta from the fixed-target collision using {\tt PYTHIA8}, and then, as an approximation of focussing, modify their direction so they are moving along the $\hat z$ axis before decaying into neutrinos. Neutrinos from negatively charged pions are ignored.  This estimate is meant for illustrative purposes only.  It overestimates both the focussing of right-sign pions and the defocussing of wrong-sign pions.  We do not use the neutrino information depicted in Fig.~\ref{fig:ETheta} (right) elsewhere in this manuscript.  Instead, when we discuss background estimates, we use the NuMI neutrino flux at MicroBooNE reported by the collaboration~\cite{neutrinoFlux@MB}.

As expected, the dark matter beam is less focused than the neutrino beam. The insert in Fig.~\ref{fig:ETheta} depicts the different angular distributions (the energy integral of the corresponding double-differential distributions) for $\chi$ and for neutrinos, where the neutrino differential distribution is multiplied by $10^{-9}$ so it fits in the same display window. 
Detectors located off-axis with respect to the neutrino beam direction witness an enhanced dark matter flux to neutrino flux ratio. This is useful for suppressing the neutrino background in the dark matter detection.  Furthermore, the NuMI beam is 120 GeV compared to the 8 GeV Booster Neutrino Beam.  This allows for production of higher energy dark matter beams.


There are three liquid argon (LAr) detectors aligned with the Booster Neutrino Beam at Fermilab and off-axis relative to the higher energy NuMI beam: MicroBooNE, ICARUS and SBND. MicroBooNE is currently taking data, ICARUS is being installed and commissioned, and SBND is in the design and construction phase. MicroBooNE has already collected over $10^{21}$~POT from NuMI. By 2024, we expect each experiment to be exposed to almost $10^{22}$~POT from NuMI \cite{NuMIexposure}.  All are designed for reconstructing and measuring charged tracks and efficiently discriminating different final-state particles, including electrons, photons, muons, pions, and protons. 
 Table~\ref{tab:DetectorFacts} lists the off-axis angle, distance from the NuMI target, and active LAr mass for each detector. Their off-axis angles are also indicated in Fig.~\ref{fig:ETheta} (solid, vertical lines).
\begin{table}[t]
\begin{center}
\begin{tabular}{cccc}
\hline\hline
Detector & MicroBooNE & ICARUS & \ \ \ SBND \ \ \ \\
\hline
Off-axis angle (rad) & 0.13  & 0.1  & 0.5   \\ 
Distance from NuMI target (m) & 684 & 789 & 409  \\ 
Active LAr mass (ton) & 89 & 476 & 112 \\
$\chi$ rate $\times$ detector mass (MicroBooNE units)  & 1 & 6.2 & 0.1 \\
\hline\hline
\end{tabular}
\end{center}
\caption{Parameters of the three Fermilab SBN liquid argon detectors with respect to the NuMI beamline (120 GeV protons).}\label{tab:DetectorFacts}
\end{table}

The number of $\chi$ particles that reaches a point-like detector from the NuMI target is 
\begin{equation}
\label{eq:fluxdist}
N_{\chi\mathrm{@det}} \simeq  \frac{A_{\mathrm{det}}}{ R_{\mathrm{det}}^2} 
\left.  \frac{dN_\chi}{d\Omega_\chi}\right|_\mathrm{det} \ , 
\end{equation}
where $R_\mathrm{det}$ is the distance to the detector and $A_\mathrm{det}$ is its cross sectional area.
For reference, in the mass range of interest, approximately $10^{-5}$ of the $\chi$ particles produced at NuMI pass through the MicroBooNE detector.
The signal event rate is proportional to the number of dark matter particles traversing the detector multiplied by the probability that it interacts in the detector volume, which is proportional to the geometric depth of the detector and the number density of scatterers. The area in Eq.~(\ref{eq:fluxdist}) combines with the depth to yield a volume, which multiplied by the number density is proportional to total mass. The figure of merit for the rate of any signal is thus proportional to the $\chi$ production rate at the detector's angle times the detector mass. Table~\ref{tab:DetectorFacts} also lists this figure of merit for each detector, normalized to that of MicroBooNE. ICARUS has a relative advantage due to its larger mass while SNBD sees a smaller dark matter flux because of the relatively large off-axis angle. In the following section, we focus on the dark trident signal, its rate, and its various properties that are relevant for all of these detectors. A detailed understanding of the idiosyncrasies of each individual experiment, required in order to properly compare the sensitivity of the different LAr detectors along the Fermilab Booster Beam, is beyond the scope of this manuscript.

\section{Dark Trident Signal}
\label{sec:signal}

The incoming beam of relativistic dark matter can scatter off an argon nucleus via a $t$-channel dark photon exchange, mimicking the neutral current interaction of a neutrino with argon.  At higher order in the dark gauge coupling $g_D$, there are processes where this scatter is accompanied with additional dark photon emission.  We consider here the case where a single dark photon is radiated off the initial or final-state $\chi$ and leave the discussion of multiple emissions to Section~\ref{sec:multi}.  Since we consider the regime $M_{A'}<2M_\chi$, the emitted dark photon(s) will decay to $e^\pm$ or $\mu^\pm$ pairs. The case of single $A'$ emission the final state is a dark trident, the corresponding Feynman diagram depicted in Fig.~\ref{fig:proddet} (right).  In what follows we will use three benchmarks to illustrate kinematic distributions and efficiency rates.  We consider $M_{A'} = 100\,\MeV,50\,\MeV,10\,\MeV$, and $M_\chi/M_{A'}=0.6$.  The results are similar for other values of $M_\chi/M_{A'}$.  Although the lightest $M_{A'}$ point is already ruled out, we believe it is still useful in order to estimate how the results evolve with mass.

\begin{figure}[t]
\centering
\includegraphics[width=0.48\textwidth]{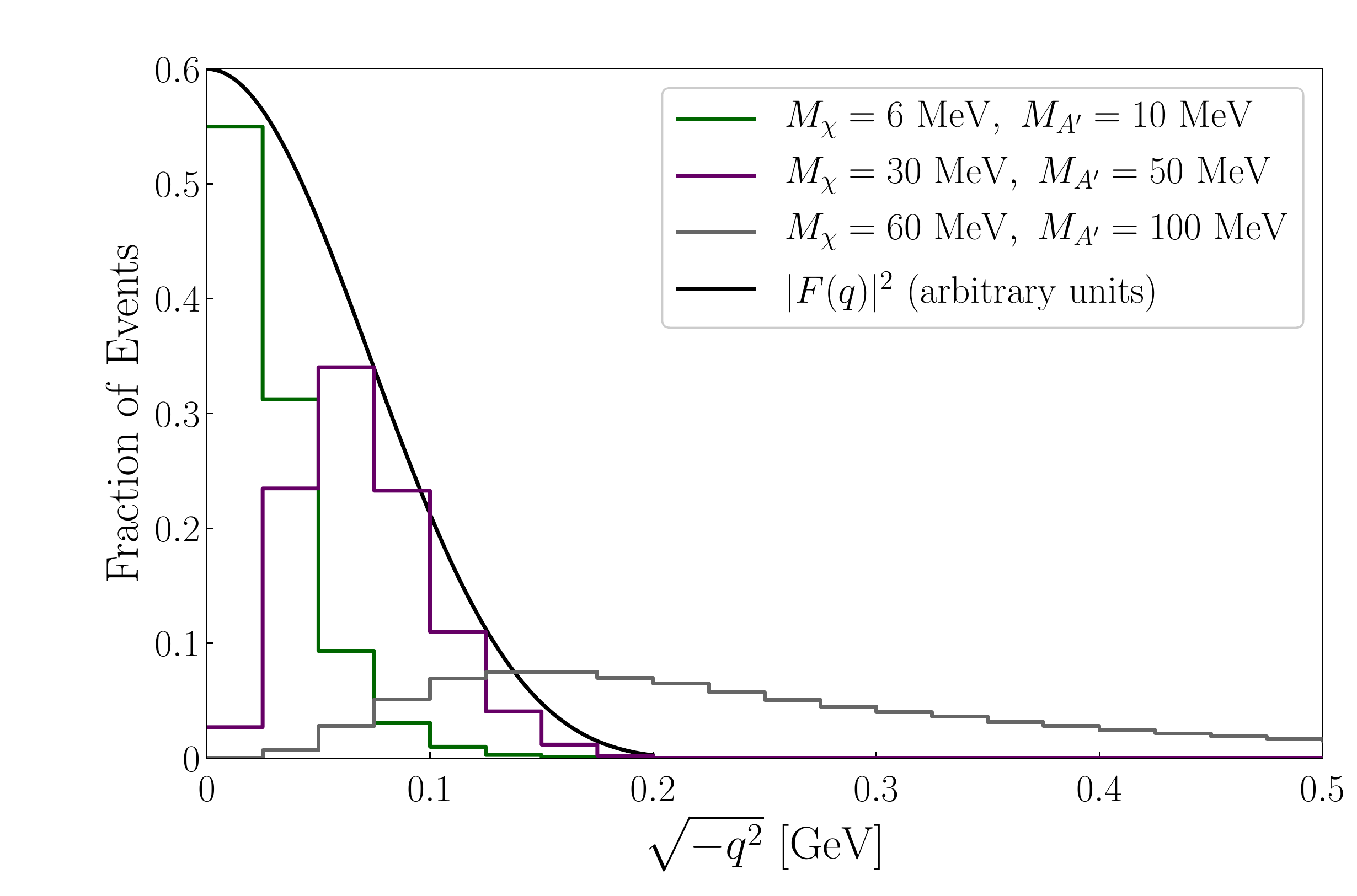}
\includegraphics[width=0.48\textwidth]{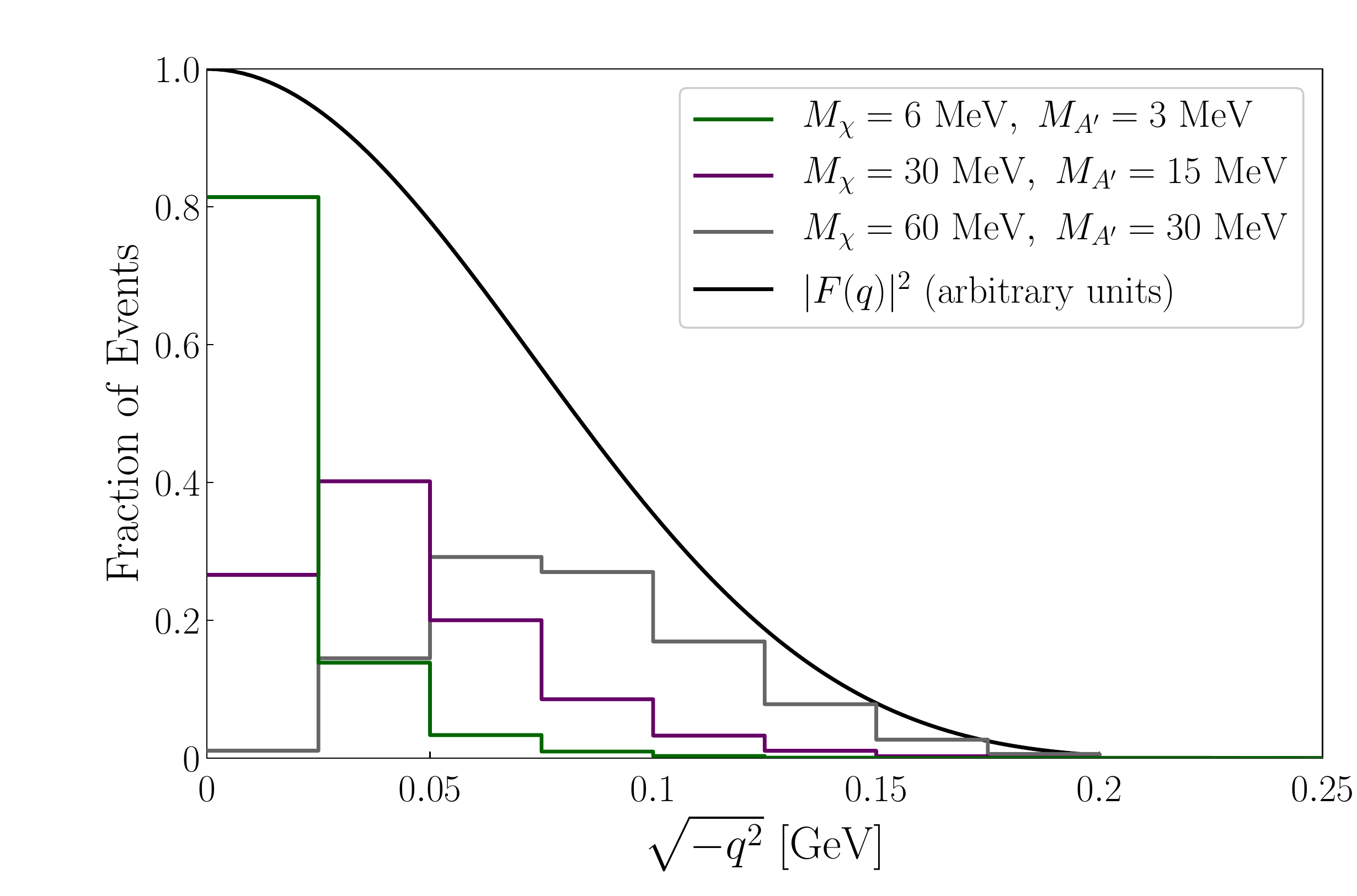}
\caption{The distribution of the momentum transfer to the argon nucleus for six dark matter and dark photon mass combinations. 
The dark matter to dark photon mass ratio is set to $0.6$ (2) in the left (right) plot.
The black curves in the same plots show the Helm form factor $F(q^2)$. In the left plot $F(q^2)$ has been scaled by $0.6$ for clarity.}\label{fig:q2}
\end{figure}

For a light $A'$, the scattering on argon is largely coherent and is enhanced by a factor of $Z_{Ar}^2\sim 300$.  However, as the magnitude of the exchanged momentum $q\equiv\sqrt{|q^2|}$ increases, coherence is lost.  This is accounted for through the inclusion of a form factor, $F(q^2)$, and here we adopt the Helm form factor \cite{Lewin:1995rx}
\be
F(q^2) = \frac{3 j_1(q R_1)}{q R_1} e^{-q^2 s^2/2} \ ,\label{eq:Helm}
\ee
where $j_1$ is the spherical Bessel function of the first kind, $R_1=\sqrt{c^2+7\pi^2 a^2/3 - 5s^2}$, $c=(1.23 A^{1/3} -0.6)\,{\rm fm}$, $s=0.9\,$fm, and $a=0.52\,$fm.
The atomic number of the target argon nucleus is $A=40$.  Numerically, we find that this form factor suppression becomes significant for $q\gtrsim 150\,$MeV, see Figure~\ref{fig:q2}.  With the inclusion of the form factor, the total event rate is given by the convolution of the scattering cross section for the process $\chi +\mathrm{Ar}\to \chi+\mathrm{Ar}+A^\prime$, $\sigma(E_\chi)$, for fixed incoming $\chi$ energy, $E_\chi$, and the number of incoming DM particles at each energy, Eq.~(\ref{eq:fluxdist}), 
\begin{equation}
N_{\rm signal} = \frac{M_{\mathrm{det}}}{R_{\mathrm{det}}^2} \int dE_\chi \int dq^2 \left|F(q^2)\right|^2 \left. \frac{d^2 N_\chi}{dE_\chi d\Omega_\chi}\right|_{\mathrm{det}}
\frac{d\sigma(E_\chi)}{dq^2}~.
\end{equation}

We have assumed the detector is effectively point like so the total rate depends upon the angular location, $\Omega_{\mathrm{det}}$, and its total mass, $M_{\mathrm{det}}=n_{\mathrm{Ar}} A_{\rm det} L_{\rm det}$, but not its orientation.
As can be seen from Figure~\ref{fig:proddet}, the number of signal events is proportional to 
\begin{equation}
N_{\rm signal} \propto \varepsilon^4 \alpha_D^3 \ .
\end{equation}

The dark trident signal arises when the final state $A'$ decays into $e^+e^-$ (or $\mu^+\mu^-$) pairs. Henceforth, because we are mostly interested in $A'$ masses around or below the $\mu^+\mu^-$ threshold, we will consider exclusively the electronic channel, and assume the branching ratio for $A'\to e^+e^-$ is one hundred percent. 
For the parameter space relevant to this work, the decay of $A'$ is prompt. The $A'$ produced through scattering with argon is boosted in the lab frame so the resulting $e^+e^-$ are collimated.  The four momenta of the outgoing lepton pair reconstruct the $A'$ mass.  Thus, ignoring the electron mass, the lepton energies, $E_{e^\pm}$, and opening angle, $\Delta\theta$, always satisfy the kinematical relation,
\be\label{eq:resonance}
4 E_{e^+} E_{e^-} \sin^2\left(\frac{\Delta\theta}{2}\right) = M_{A'}^2  \ .
\ee
The existence of a resonance and the different kinematics of the dark matter signal and the neutrino background will allow separation of signal from background events.  We address the issues of background rate and signal efficiency below, in Section~\ref{sec:signalefficiencyandbackground}, but point out here that the true reach can only be determined by each experiment.  Here we present the best case scenario, the region of the $\varepsilon^2-M_{A'}$ parameter space for which MicroBooNE, ICARUS, and SBND will have 10 or more dark trident signal events after a fixed amount of POT in the NuMI beam, assuming the efficiency for the signal is 100\%. We consider two points in time -- the present and the mid-20's.  The MicroBooNE experiment has already collected over $10^{21}$~POT from the NuMI beam. The region of parameter space where MicroBooNE expected to have already recorded ten signal events assuming maximum signal efficiency is depicted by the dashed black line in Fig.~\ref{fig:MicroBooNERegions} and Fig.~\ref{fig:MicroBooNERegionsADM}. On the other hand, the NuMI beam is expected to run until 2024 and it is not overly optimistic to assume it will collect over $8\times 10^{21}$~POT. The region of parameter space where the three different LAr experiments are expected to record ten signal events assuming maximum signal efficiency is depicted by the solid lines in Fig.~\ref{fig:MicroBooNERegions} and Fig.~\ref{fig:MicroBooNERegionsADM}, the color coding is indicated in the figure. As we will argue below, (our estimate of) the efficiency is high for $M_{A'}\gsim 50$~MeV but for lower masses the efficiency drops and one would require more POTs to identify the same number of events. Taking efficiency into account, the curves in Figs.~\ref{fig:MicroBooNERegions}, \ref{fig:MicroBooNERegionsADM} can be thought of as the contours for 10 dark trident signal events with luminosity rescaled by 1/efficiency. We emphasize that efficiencies and background rates will vary by experiment and can only be reliably determined by them. Furthermore, the ultimate results presented in Figure~\ref{fig:MicroBooNERegions}, \ref{fig:MicroBooNERegionsADM} assume equal exposure for all three experiments.

\subsection{Estimates of Background Rates and Signal Efficiency}\label{sec:signalefficiencyandbackground}

The irreducible background for our signal is SM trident production from the off-axis NuMI neutrino beam passing through the MicroBooNE detector.  The SM neutrino trident production cross section is known to be very small~\cite{Altmannshofer:2014pba,Ge:2017poy}, $\sigma\left(E_\nu\lsim 10\,\mathrm{GeV}\right)\sim 10^{-6}\,\mathrm{pb}$.  Using the officially reported NuMI neutrino flux at MicroBooNE~\cite{neutrinoFlux@MB} we estimate that there is less than a $1\%$ chance of one neutrino trident event for $10^{21}$ POT.  Instead, the dominant background is reducible and comes from the normal neutrino charged current (CC) or neutral-current (NC) interactions dressed with additional $\pi^0$ radiation.  These dressed events could fake the trident signal if photons are misidentified as electrons.  In addition, $\gamma$ emission can also fake the signal.  The associated rate is sufficiently small and can be ignored.

Based on the officially reported NuMI neutrino flux at MicroBooNE~\cite{neutrinoFlux@MB} and the ArgoNeuT measurement of the CC scattering cross section on argon~\cite{Acciarri:2014isz}, we estimate $\sim 6\times 10^4 (3000)$ $\nu_\mu(\nu_e)$ CC events at MicroBooNE, for $10^{21}$ POT.  The $\nu_e$ CC events can fake the signal if dressed by a $\pi^0$, the rate for which was measured at MINERvA~\cite{Aliaga:2015wva}.  We expect $\sim 600$ events.  The $\nu_\mu$ CC events will not fake the signal unless a muon is misidentified as an electron, and the event is dressed with a $\pi^0$.  However, $\nu_\mu$ NC events need only be dressed with a $\pi^0$ (NC$\pi^0$) to fake the signal.  The ratio of this rate to the CC rate was measured at ArgoNeuT \cite{Acciarri:2015ncl} and is $\sigma_{NC\pi^0}/\sigma_{CC}\sim 0.14$.  Thus, we expect $\sim 8000$ such events.
Liquid argon detectors have the capability of precise particle identification. Leveraging features like energy deposition and the decay length of photons in liquid argon~\cite{Caratelli:2018nob}, we expect low misidentification rates, such as $\sim 10^{-2}$ for confusing a $\pi^0$ with an electron or $10^{-3}$ for confusing a $\pi^0$ with an electron pair. With this level of background rejection, we expect a total of $\mathcal{O}(15)$ background events, before taking into account the invariant mass peak of the signal we are searching for, which is lacking in the background.

It is interesting to compare the signal and background rates for an off-axis detector like MicroBooNE to those for a more on-axis detector like NOvA, or in the future DUNE\footnote{The DUNE-PRISM proposal \cite{duneprism} for a moveable DUNE near detector to reduce systematic effects in determining oscillation parameters will function as both an on-axis and off-axis detector, and will be a good dark trident detector.}.  The signal rate for a detector at the same off-axis angle as NOvA (14 mrad) is larger than that at the angle of MicroBooNE by a factor of 20.  On the other hand, backgrounds are roughly 200 times larger.  Thus, the search for dark tridents changes from being a statistics-limited analysis to being a systematics-limited one.  Therefore, one must understand the background rate more precisely than we have.

\begin{table}[t]
\begin{center}
\begin{tabular}{cccc||cccc}
\hline\hline
\multicolumn{4}{c||}{$E_\mathrm{min}, E_\mathrm{max} > 30$ MeV} & \multicolumn{4}{c}{$E_\mathrm{max} > 30$ MeV, $E_\mathrm{min} > 5$ MeV} \\
\hline
\quad \quad \quad \quad  $M_{A^\prime}/M_\chi$ \quad \quad \quad \quad & 5$^\circ$ & 8$^\circ$ & 15$^\circ$ & \quad \quad \quad \quad $M_{A^\prime}/M_\chi$ \quad \quad \quad \quad & 5$^\circ$ & 8$^\circ$ & 15$^\circ$ \\ \hline
100 MeV/60 MeV & 0.75\ \  & 0.70\ \   & 0.55\ \   & 100 MeV/60 MeV & 0.95\ \   & 0.90\ \   & 0.75\ \   \\
50 MeV/30 MeV & 0.56\ \  & 0.44\ \   & 0.26\ \   & 50 MeV/30 MeV & 0.79\ \   & 0.68\ \   & 0.49\ \   \\
10 MeV/6 MeV & 0.11\ \  & 0.028\ \   & 0.001\ \   & 10 MeV/6 MeV & 0.29\ \   & 0.15\ \   & 0.04\ \   \\ 
\hline\hline
\end{tabular}
\end{center}
\caption{Efficiencies to pass cuts on the $e^\pm$ pair produced in DM trident events, estimated from Monte Carlo.}\label{tab:efficiencies}
\end{table}

\begin{figure}[t]
\centering
\includegraphics[width=0.9\textwidth]{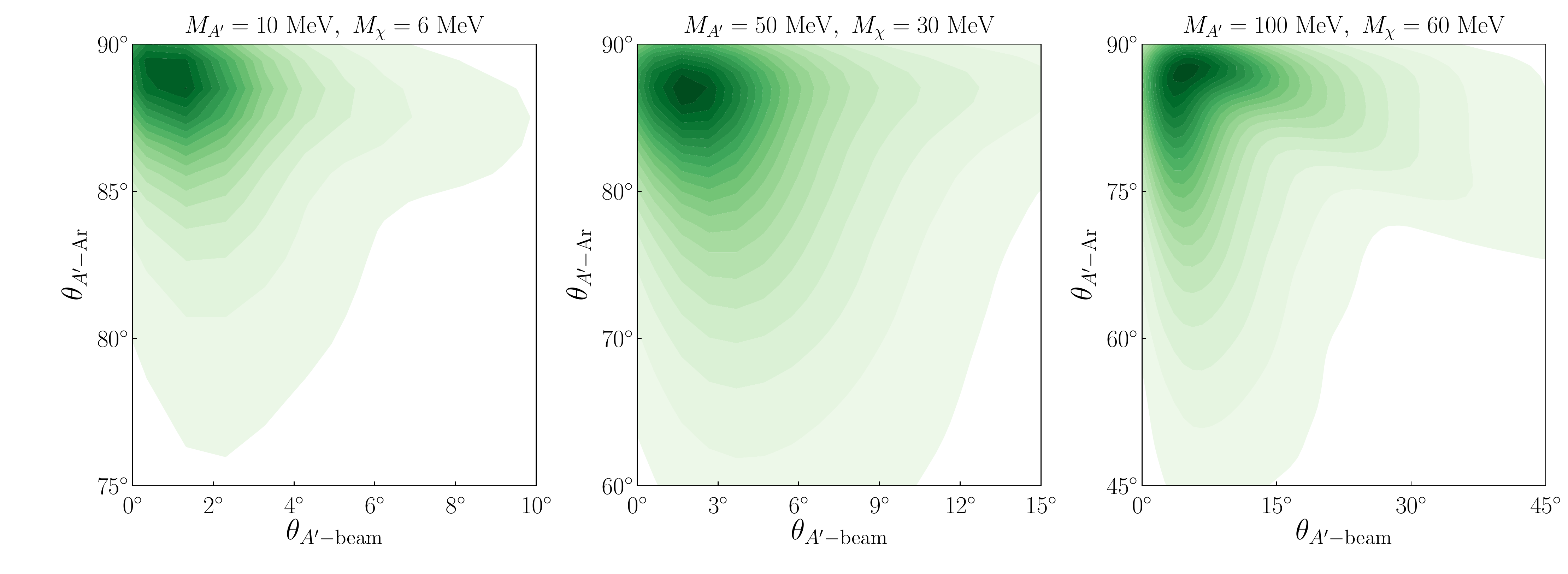}
\caption{The angular distributions of the final state $A'$ with respect to the incoming dark matter beam direction ($\theta_{A'-\text{beam}}$) and with respect to the recoiling argon nucleus direction ($\theta_{A'-Ar}$), for three dark matter and dark photon mass combinations. 
We fix the dark matter to dark photon mass ratio to be $M_\chi/M_{A'}=0.6$. Note the different ranges in the axes of the different panels.
}\label{fig:theta2theta}
\end{figure}

The DM trident signal is a rich final state providing many observables that LAr detectors can measure precisely, allowing signal to be separated from backgrounds.  The two electron tracks reconstruct a resonance of fixed invariant mass Eq.~(\ref{eq:resonance}), the reconstructed $A'$ momentum points back to the target, and the arrival of the electrons is in time with the beam.  There may also be additional activity from the recoil of the argon nucleus that can potentially be observed and is correlated with the electron pair.  In particular, the dark photon, reconstructed from the electron pair, tends to travel along the beam direction while the argon nucleus is mostly likely to travel in the orthogonal direction to the dark photon. These correlations are demonstrated in Fig.~\ref{fig:theta2theta}.  For light mediators ($M_{A'}\lsim 100$ MeV), the energy transfer to the argon nucleus is typically well below its binding energy and the hadronic activity will be small.  In this case, we may further suppress the background by requiring very little hadronic activity in the events. Other cuts, such as requiring the charged leptons to have nearly the same track length, may also aid in reducing background events.

\begin{figure}[t]
\centering
\includegraphics[width=0.9\textwidth]{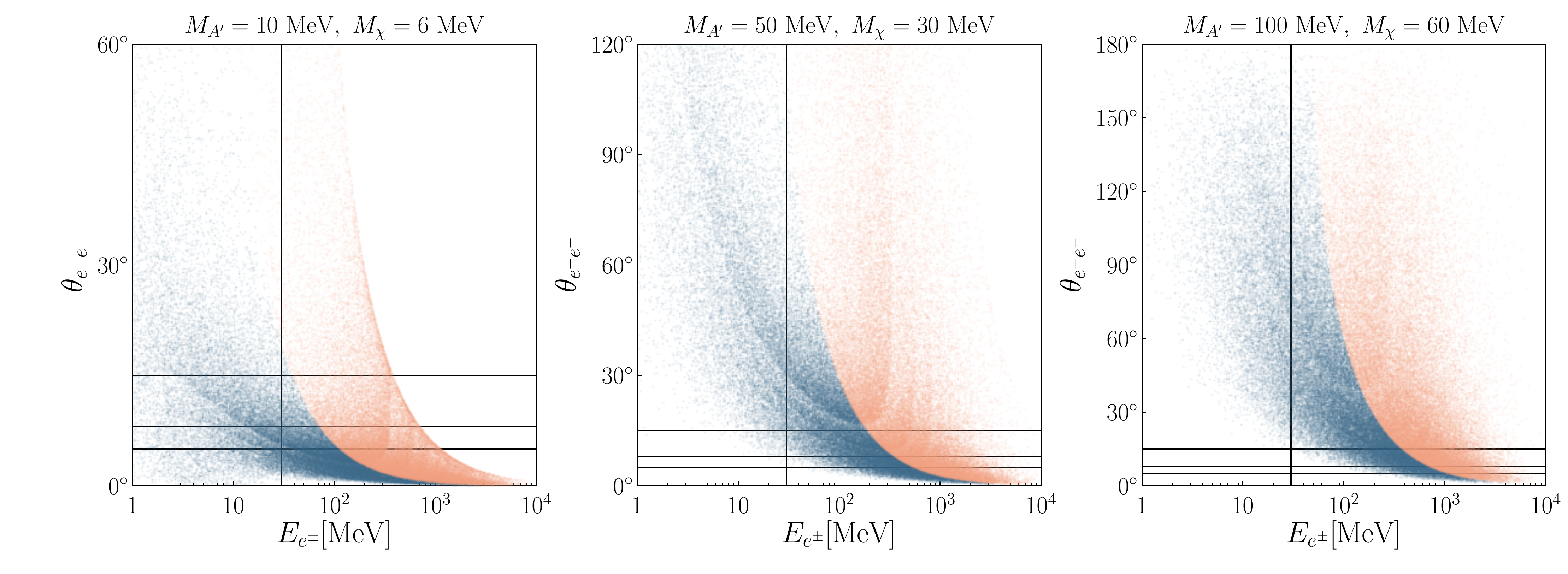}
\caption{The final state $e^+e^-$ opening angle versus energy distribution for three dark matter and dark photon mass combinations. 
We fix the dark matter to dark photon mass ratio to be $0.6$. 
In each event, the energy of the more (less) energetic charged lepton is represented by an orange (blue) point. 
The distributions are obtained by convolving with the energy spectrum of the incoming $\chi$ beam.
The patterns in the bulk of each region are numerical artifacts.
The vertical line corresponds to $E_{e^\pm} = 30\,$MeV.
The horizontal lines correspond to $\theta_{e^+e^-}= 5^\circ,8^\circ,15^\circ$.}\label{fig:theta2E}
\end{figure}

Since the $e^\pm$ pair reconstructs a resonance, their energies and the opening angle between them are related, as discussed in Eq.~(\ref{eq:resonance}) and depicted in Fig.~\ref{fig:theta2E}.  There are minimum energy requirements for the LAr detectors to accurately measure the electrons' momentum and energy and the lepton pair must be sufficiently separated to be identified as two objects.  Exactly what these requirements are depends on complicated detector issues, but it is clear from Eq.~(\ref{eq:resonance}) and Fig.~\ref{fig:theta2E} that lighter $A'$ decays will be harder to identify as the signal.  Rather than attempt a detailed detector simulation, we investigate the effects of some reasonable requirements on the lepton energies and $\Delta\theta$.  In Table~\ref{tab:efficiencies}, we list the efficiency for the events to pass a series of requirements on the leptons for a few benchmarks.  We consider two choices for energy cuts: a conservative choice, which requires both leptons to have $E>30$ MeV, and a more aggressive one, which requires only the leading lepton energy to be above 30 MeV and that the other be above 5 MeV.  In each case we consider requiring separation between the electrons of 5$^\circ$, 8$^\circ$, or 15$^\circ$; these various cuts are depicted as lines in Fig.~\ref{fig:theta2E}.  The heavier $A'$ benchmark passes all choices of cuts with high efficiency, whereas the lightest $A'$ benchmark only has reasonable efficiency for the more aggressive choice, where small angular separations are resolved.

\section{Beyond Trident: Multiple Dark Photon Radiations}
\label{sec:multi}

Here, we comment on the possibility of ``multi-trident'' dark matter production, where the incident $\chi$ flux generates the process $\chi + \mathrm{Ar} \to \chi + \mathrm{Ar} + nA^\prime$, where $n\geq 2$. The $n$ on-shell $A^\prime$ particles then decay to electron/positron pairs, producing a signal that is even more striking than a single dark trident event. This process's cross section depends on two features: first, $M_{A^\prime}$ must be light enough to be radiated on-shell; secondly, the coupling $g_D$ must be relatively large.
\begin{table}[t]
\begin{center}
\begin{tabular}{ccc}\hline\hline
$M_{A^\prime}$ (MeV)\ \  & $M_\chi$ (MeV) \ \ & $N_{n=2} / N_{n=1}$  \ \ \\ \hline 
100 & 60 & $0.06\,\alpha_D$ \\ 
50 & 30 & $0.14\,\alpha_D$ \\ 
10 & 6 & $0.38\,\alpha_D$ \\ 
\hline\hline 
\end{tabular}
\end{center}
\caption{The number of events at MicroBooNE for double-$A^\prime$ emission relative to single-$A^\prime$ emission, for three sets of $\chi$ and $A^\prime$ masses.}
\label{Table:DoubleEmission}
\end{table}
In Table~\ref{Table:DoubleEmission}, we show the relative number of events for $n=2$ \emph{vs.} $n=1$ for three different combinations of $M_{A^\prime}$ and $M_\chi$. We see that the $n=2$ cross section is more relevant for lighter $A^\prime$, and can be significant if $\alpha_D \sim 1$. This is opposite to the effects of reconstruction efficiency discussed above: while lighter $A^\prime$ may be harder to detect in the single-trident channel, the abundance of multi-trident events could make these searches more feasible.

The possibility of multi-trident events leads to interesting complications for event reconstruction. As discussed, see Fig.~\ref{fig:theta2E}, for boosted $A^\prime$ decaying to $e^+ e^-$, the electron and positron can have a small opening angle, making it difficult to identify a signal event. This problem is further complicated in multi-trident events, and there is a possibility of four charged leptons being identified as 1, 2, 3, or 4 independent tracks. Additionally, we discussed using the invariant mass of the charged lepton pair as a way to reduce neutrino-related backgrounds. With multi-trident events, combinatorics makes it more difficult to identify which tracks originated from the same $A^\prime$, complicating this reconstruction.  We leave an analysis of this more complicated final state to future work.

\section{Conclusion and Outlook}
\label{sec:conclusions}

In this work, we explore a new channel for neutrino experiments to probe the dark sector. The dark trident signal consists of a resonant dilepton pair from the prompt decay of a dark photon, which is produced in the collision of a dark state with a nucleus.
We consider a simple model where the dark matter interacts through the dark photon portal. Our study focuses on the parameter space where the dark photon cannot decay into dark matter particles and visible decays dominate. 
In this case, a dark matter beam can be created in fixed-target collisions through the production of off-shell dark photons. The dark tridents are then created downstream in DM-nucleus collisions. 

Within this model, we have shown that the dark matter beam is broader than the neutrino beam. As a result, the ratio of dark trident signal to neutrino backgrounds is larger for detectors that are a few degrees off-axis.
We thus consider dark matter produced at the NuMI target at Fermilab and then travels to the off-axis liquid argon detectors, MicroBooNE, ICARUS and SBND. This allows one to search for dark tridents with low background rates and to probe currently unexplored regions of parameter space, as depicted in Figs.~\ref{fig:MicroBooNERegions} and \ref{fig:MicroBooNERegionsADM}. On-axis detectors may also probe this dark sector model, but require a more  precise understanding of the background when compared to the off-axis case. One can also envision dark trident signals induced by new physics in neutrino scattering~(e.g.~\cite{Bertuzzo:2018itn, Ballett:2018ynz}). In these cases, on-axis detectors would preform better than off-axis ones.

We have studied the expected efficiency for dark tridents based on the planned capabilities of LAr detectors and found interesting complementary effects. When the dark photon is relatively heavy, of order 100~MeV, the opening angle between the daughter electron and positron is sizable and the expected reconstruction efficiency is high. In the region of lighter $A'$, of order 10~MeV, the leptons are more collinear and harder to reconstruct. However, in this region of parameter space, the signal rate is expected to be higher. In addition, we find that in this lighter $A'$ region the rate for multi-$A'$ production, which is likely a more distinct signal, is higher. 

The current and upcoming SBN program at Fermilab has a promising opportunity to investigate dark sectors in the dark trident channel. The analysis of MicroBooNE data currently on tape is already capable of probing unexplored regions of the parameter space.

\subsection*{Acknowledgments}
We thank Brian Batell, Virgil Bocean, David Caratelli, Raquel Castillo, Joshua Isaacson, Gordan Krnjaic, Ornella Palamara, and Yun-Tse Tsai.
PF, RH, and YZ would like to thank the Aspen Center for Physics, which is supported by National Science Foundation grant PHY-1607611, where part of this work was performed. KJK thanks the Fermilab Neutrino Physics Center for support during the work of this manuscript.  YZ thanks Colegio De Fisica Fundamental E Interdisciplinaria De Las Americas (COFI) for travel support during the completion of this work.  AdG, KJK and YZ are supported in part by DOE grant \#de-sc0010143.  The work of PF, RH, KJK and YZ was supported in part by the DoE under contract number DE-SC0007859 and Fermilab, operated by Fermi Research Alliance, LLC under contract number DE-AC02-07CH11359 with the United States Department of Energy.  


\bibliographystyle{JHEP}
\bibliography{References}

\providecommand{\href}[2]{#2}\begingroup\raggedright\begin{thebibliography}{10}

\bibitem{Batell:2009di}
B.~Batell, M.~Pospelov and A.~Ritz, \emph{{Exploring Portals to a Hidden Sector
  Through Fixed Targets}},
  \href{http://dx.doi.org/10.1103/PhysRevD.80.095024}{\emph{Phys. Rev.} {\bf
  D80} (2009) 095024}, [\href{http://arxiv.org/abs/0906.5614}{{\tt
  0906.5614}}].

\bibitem{Essig:2010gu}
R.~Essig, R.~Harnik, J.~Kaplan and N.~Toro, \emph{{Discovering New Light States
  at Neutrino Experiments}},
  \href{http://dx.doi.org/10.1103/PhysRevD.82.113008}{\emph{Phys. Rev.} {\bf
  D82} (2010) 113008}, [\href{http://arxiv.org/abs/1008.0636}{{\tt
  1008.0636}}].

\bibitem{deNiverville:2011it}
P.~deNiverville, M.~Pospelov and A.~Ritz, \emph{{Observing a light dark matter
  beam with neutrino experiments}},
  \href{http://dx.doi.org/10.1103/PhysRevD.84.075020}{\emph{Phys. Rev.} {\bf
  D84} (2011) 075020}, [\href{http://arxiv.org/abs/1107.4580}{{\tt
  1107.4580}}].

\bibitem{Dobrescu:2014ita}
B.~A. Dobrescu and C.~Frugiuele, \emph{{GeV-Scale Dark Matter: Production at
  the Main Injector}},
  \href{http://dx.doi.org/10.1007/JHEP02(2015)019}{\emph{JHEP} {\bf 02} (2015)
  019}, [\href{http://arxiv.org/abs/1410.1566}{{\tt 1410.1566}}].

\bibitem{Coloma:2015pih}
P.~Coloma, B.~A. Dobrescu, C.~Frugiuele and R.~Harnik, \emph{{Dark matter beams
  at LBNF}}, \href{http://dx.doi.org/10.1007/JHEP04(2016)047}{\emph{JHEP} {\bf
  04} (2016) 047}, [\href{http://arxiv.org/abs/1512.03852}{{\tt 1512.03852}}].

\bibitem{Aguilar-Arevalo:2017mqx}
{\scshape MiniBooNE} collaboration, A.~A. Aguilar-Arevalo et~al., \emph{{Dark
  Matter Search in a Proton Beam Dump with MiniBooNE}},
  \href{http://dx.doi.org/10.1103/PhysRevLett.118.221803}{\emph{Phys. Rev.
  Lett.} {\bf 118} (2017) 221803}, [\href{http://arxiv.org/abs/1702.02688}{{\tt
  1702.02688}}].

\bibitem{Frugiuele:2017zvx}
C.~Frugiuele, \emph{{Probing sub-GeV dark sectors via high energy proton beams
  at LBNF/DUNE and MiniBooNE}},
  \href{http://dx.doi.org/10.1103/PhysRevD.96.015029}{\emph{Phys. Rev.} {\bf
  D96} (2017) 015029}, [\href{http://arxiv.org/abs/1701.05464}{{\tt
  1701.05464}}].

\bibitem{Izaguirre:2017bqb}
E.~Izaguirre, Y.~Kahn, G.~Krnjaic and M.~Moschella, \emph{{Testing Light Dark
  Matter Coannihilation With Fixed-Target Experiments}},
  \href{http://dx.doi.org/10.1103/PhysRevD.96.055007}{\emph{Phys. Rev.} {\bf
  D96} (2017) 055007}, [\href{http://arxiv.org/abs/1703.06881}{{\tt
  1703.06881}}].

\bibitem{Magill:2018jla}
G.~Magill, R.~Plestid, M.~Pospelov and Y.-D. Tsai, \emph{{Dipole portal to
  heavy neutral leptons}},  \href{http://arxiv.org/abs/1803.03262}{{\tt
  1803.03262}}.

\bibitem{Magill:2018tbb}
G.~Magill, R.~Plestid, M.~Pospelov and Y.-D. Tsai, \emph{{Millicharged
  particles in neutrino experiments}},
  \href{http://arxiv.org/abs/1806.03310}{{\tt 1806.03310}}.

\bibitem{deNiverville:2018dbu}
P.~deNiverville and C.~Frugiuele, \emph{{Hunting sub-GeV dark matter with
  NO$\nu$A near detector}},  \href{http://arxiv.org/abs/1807.06501}{{\tt
  1807.06501}}.

\bibitem{Bjorken:2009mm}
J.~D. Bjorken, R.~Essig, P.~Schuster and N.~Toro, \emph{{New Fixed-Target
  Experiments to Search for Dark Gauge Forces}},
  \href{http://dx.doi.org/10.1103/PhysRevD.80.075018}{\emph{Phys. Rev.} {\bf
  D80} (2009) 075018}, [\href{http://arxiv.org/abs/0906.0580}{{\tt
  0906.0580}}].

\bibitem{Abrahamyan:2011gv}
{\scshape APEX} collaboration, S.~Abrahamyan et~al., \emph{{Search for a New
  Gauge Boson in Electron-Nucleus Fixed-Target Scattering by the APEX
  Experiment}},
  \href{http://dx.doi.org/10.1103/PhysRevLett.107.191804}{\emph{Phys. Rev.
  Lett.} {\bf 107} (2011) 191804}, [\href{http://arxiv.org/abs/1108.2750}{{\tt
  1108.2750}}].

\bibitem{Battaglieri:2014hga}
M.~Battaglieri et~al., \emph{{The Heavy Photon Search Test Detector}},
  \href{http://dx.doi.org/10.1016/j.nima.2014.12.017}{\emph{Nucl. Instrum.
  Meth.} {\bf A777} (2015) 91--101},
  [\href{http://arxiv.org/abs/1406.6115}{{\tt 1406.6115}}].

\bibitem{Berlin:2018pwi}
A.~Berlin, S.~Gori, P.~Schuster and N.~Toro, \emph{{Dark Sectors at the
  Fermilab SeaQuest Experiment}},  \href{http://arxiv.org/abs/1804.00661}{{\tt
  1804.00661}}.

\bibitem{Pospelov:2007mp}
M.~Pospelov, A.~Ritz and M.~B. Voloshin, \emph{{Secluded WIMP Dark Matter}},
  \href{http://dx.doi.org/10.1016/j.physletb.2008.02.052}{\emph{Phys. Lett.}
  {\bf B662} (2008) 53--61}, [\href{http://arxiv.org/abs/0711.4866}{{\tt
  0711.4866}}].

\bibitem{Feng:2008ya}
J.~L. Feng and J.~Kumar, \emph{{The WIMPless Miracle: Dark-Matter Particles
  without Weak-Scale Masses or Weak Interactions}},
  \href{http://dx.doi.org/10.1103/PhysRevLett.101.231301}{\emph{Phys. Rev.
  Lett.} {\bf 101} (2008) 231301}, [\href{http://arxiv.org/abs/0803.4196}{{\tt
  0803.4196}}].

\bibitem{Pospelov:2008jd}
M.~Pospelov and A.~Ritz, \emph{{Astrophysical Signatures of Secluded Dark
  Matter}}, \href{http://dx.doi.org/10.1016/j.physletb.2008.12.012}{\emph{Phys.
  Lett.} {\bf B671} (2009) 391--397},
  [\href{http://arxiv.org/abs/0810.1502}{{\tt 0810.1502}}].

\bibitem{ArkaniHamed:2008qn}
N.~Arkani-Hamed, D.~P. Finkbeiner, T.~R. Slatyer and N.~Weiner, \emph{{A Theory
  of Dark Matter}},
  \href{http://dx.doi.org/10.1103/PhysRevD.79.015014}{\emph{Phys. Rev.} {\bf
  D79} (2009) 015014}, [\href{http://arxiv.org/abs/0810.0713}{{\tt
  0810.0713}}].

\bibitem{Rocha:2012jg}
M.~Rocha, A.~H.~G. Peter, J.~S. Bullock, M.~Kaplinghat, S.~Garrison-Kimmel,
  J.~Onorbe et~al., \emph{{Cosmological Simulations with Self-Interacting Dark
  Matter I: Constant Density Cores and Substructure}},
  \href{http://dx.doi.org/10.1093/mnras/sts514}{\emph{Mon. Not. Roy. Astron.
  Soc.} {\bf 430} (2013) 81--104}, [\href{http://arxiv.org/abs/1208.3025}{{\tt
  1208.3025}}].

\bibitem{Bertuzzo:2018itn}
E.~Bertuzzo, S.~Jana, P.~A.~N. Machado and R.~Zukanovich~Funchal, \emph{{A Dark
  Neutrino Portal to Explain MiniBooNE}},
  \href{http://arxiv.org/abs/1807.09877}{{\tt 1807.09877}}.

\bibitem{Ballett:2018ynz}
P.~Ballett, S.~Pascoli and M.~Ross-Lonergan, \emph{{U(1)' mediated decays of
  heavy sterile neutrinos in MiniBooNE}},
  \href{http://arxiv.org/abs/1808.02915}{{\tt 1808.02915}}.

\bibitem{Batell:2014mga}
B.~Batell, R.~Essig and Z.~Surujon, \emph{{Strong Constraints on Sub-GeV Dark
  Sectors from SLAC Beam Dump E137}},
  \href{http://dx.doi.org/10.1103/PhysRevLett.113.171802}{\emph{Phys. Rev.
  Lett.} {\bf 113} (2014) 171802}, [\href{http://arxiv.org/abs/1406.2698}{{\tt
  1406.2698}}].

\bibitem{Kahn:2014sra}
Y.~Kahn, G.~Krnjaic, J.~Thaler and M.~Toups, \emph{{DAE${\delta}$ALUS and dark
  matter detection}},
  \href{http://dx.doi.org/10.1103/PhysRevD.91.055006}{\emph{Phys. Rev.} {\bf
  D91} (2015) 055006}, [\href{http://arxiv.org/abs/1411.1055}{{\tt
  1411.1055}}].

\bibitem{Slatyer:2015jla}
T.~R. Slatyer, \emph{{Indirect dark matter signatures in the cosmic dark ages.
  I. Generalizing the bound on s-wave dark matter annihilation from Planck
  results}}, \href{http://dx.doi.org/10.1103/PhysRevD.93.023527}{\emph{Phys.
  Rev.} {\bf D93} (2016) 023527}, [\href{http://arxiv.org/abs/1506.03811}{{\tt
  1506.03811}}].

\bibitem{Slatyer:2009yq}
T.~R. Slatyer, N.~Padmanabhan and D.~P. Finkbeiner, \emph{{CMB Constraints on
  WIMP Annihilation: Energy Absorption During the Recombination Epoch}},
  \href{http://dx.doi.org/10.1103/PhysRevD.80.043526}{\emph{Phys. Rev.} {\bf
  D80} (2009) 043526}, [\href{http://arxiv.org/abs/0906.1197}{{\tt
  0906.1197}}].

\bibitem{Madhavacheril:2013cna}
M.~S. Madhavacheril, N.~Sehgal and T.~R. Slatyer, \emph{{Current Dark Matter
  Annihilation Constraints from CMB and Low-Redshift Data}},
  \href{http://dx.doi.org/10.1103/PhysRevD.89.103508}{\emph{Phys. Rev.} {\bf
  D89} (2014) 103508}, [\href{http://arxiv.org/abs/1310.3815}{{\tt
  1310.3815}}].

\bibitem{Cline:2014dwa}
J.~M. Cline, G.~Dupuis, Z.~Liu and W.~Xue, \emph{{The windows for kinetically
  mixed Z'-mediated dark matter and the galactic center gamma ray excess}},
  \href{http://dx.doi.org/10.1007/JHEP08(2014)131}{\emph{JHEP} {\bf 08} (2014)
  131}, [\href{http://arxiv.org/abs/1405.7691}{{\tt 1405.7691}}].

\bibitem{DAgnolo:2015ujb}
R.~T. D'Agnolo and J.~T. Ruderman, \emph{{Light Dark Matter from Forbidden
  Channels}},
  \href{http://dx.doi.org/10.1103/PhysRevLett.115.061301}{\emph{Phys. Rev.
  Lett.} {\bf 115} (2015) 061301}, [\href{http://arxiv.org/abs/1505.07107}{{\tt
  1505.07107}}].

\bibitem{Cline:2017tka}
J.~M. Cline, H.~Liu, T.~Slatyer and W.~Xue, \emph{{Enabling Forbidden Dark
  Matter}}, \href{http://dx.doi.org/10.1103/PhysRevD.96.083521}{\emph{Phys.
  Rev.} {\bf D96} (2017) 083521}, [\href{http://arxiv.org/abs/1702.07716}{{\tt
  1702.07716}}].

\bibitem{Nussinov:1985xr}
S.~Nussinov, \emph{{Technocosmology: Could a Technibaryon Excess Provide a
  'Natural' Missing Mass Candidate?}},
  \href{http://dx.doi.org/10.1016/0370-2693(85)90689-6}{\emph{Phys. Lett.} {\bf
  165B} (1985) 55--58}.

\bibitem{Kaplan:2009ag}
D.~E. Kaplan, M.~A. Luty and K.~M. Zurek, \emph{{Asymmetric Dark Matter}},
  \href{http://dx.doi.org/10.1103/PhysRevD.79.115016}{\emph{Phys. Rev.} {\bf
  D79} (2009) 115016}, [\href{http://arxiv.org/abs/0901.4117}{{\tt
  0901.4117}}].

\bibitem{Essig:2012yx}
R.~Essig, A.~Manalaysay, J.~Mardon, P.~Sorensen and T.~Volansky, \emph{{First
  Direct Detection Limits on sub-GeV Dark Matter from XENON10}},
  \href{http://dx.doi.org/10.1103/PhysRevLett.109.021301}{\emph{Phys. Rev.
  Lett.} {\bf 109} (2012) 021301}, [\href{http://arxiv.org/abs/1206.2644}{{\tt
  1206.2644}}].

\bibitem{Essig:2017kqs}
R.~Essig, T.~Volansky and T.-T. Yu, \emph{{New Constraints and Prospects for
  sub-GeV Dark Matter Scattering off Electrons in Xenon}},
  \href{http://dx.doi.org/10.1103/PhysRevD.96.043017}{\emph{Phys. Rev.} {\bf
  D96} (2017) 043017}, [\href{http://arxiv.org/abs/1703.00910}{{\tt
  1703.00910}}].

\bibitem{Batley:2015lha}
{\scshape NA48/2} collaboration, J.~R. Batley et~al., \emph{{Search for the
  dark photon in $\pi^0$ decays}},
  \href{http://dx.doi.org/10.1016/j.physletb.2015.04.068}{\emph{Phys. Lett.}
  {\bf B746} (2015) 178--185}, [\href{http://arxiv.org/abs/1504.00607}{{\tt
  1504.00607}}].

\bibitem{Bjorken:1988as}
J.~D. Bjorken, S.~Ecklund, W.~R. Nelson, A.~Abashian, C.~Church, B.~Lu et~al.,
  \emph{{Search for Neutral Metastable Penetrating Particles Produced in the
  SLAC Beam Dump}},
  \href{http://dx.doi.org/10.1103/PhysRevD.38.3375}{\emph{Phys. Rev.} {\bf D38}
  (1988) 3375}.

\bibitem{Riordan:1987aw}
E.~M. Riordan et~al., \emph{{A Search for Short Lived Axions in an Electron
  Beam Dump Experiment}},
  \href{http://dx.doi.org/10.1103/PhysRevLett.59.755}{\emph{Phys. Rev. Lett.}
  {\bf 59} (1987) 755}.

\bibitem{Bross:1989mp}
A.~Bross, M.~Crisler, S.~H. Pordes, J.~Volk, S.~Errede and J.~Wrbanek, \emph{{A
  Search for Shortlived Particles Produced in an Electron Beam Dump}},
  \href{http://dx.doi.org/10.1103/PhysRevLett.67.2942}{\emph{Phys. Rev. Lett.}
  {\bf 67} (1991) 2942--2945}.

\bibitem{Lees:2014xha}
{\scshape BaBar} collaboration, J.~P. Lees et~al., \emph{{Search for a Dark
  Photon in $e^+e^-$ Collisions at BaBar}},
  \href{http://dx.doi.org/10.1103/PhysRevLett.113.201801}{\emph{Phys. Rev.
  Lett.} {\bf 113} (2014) 201801}, [\href{http://arxiv.org/abs/1406.2980}{{\tt
  1406.2980}}].

\bibitem{GordanPrivateCommunication}
{G.~Krnjaic, private communication}.

\bibitem{neutrinoFlux@MB}
{MicroBooNE Collaboration},
  ``{\url{https://microboone-exp.fnal.gov/public/approved_plots/Beam.html}}.''

\bibitem{NuMIexposure}
{NOvA Collaboration}, ``{Fermilab PAC Meeting, 2018;
  \url{https://indico.fnal.gov/event/17480/contribution/8/material/slides/1.pdf}}.''

\bibitem{Lewin:1995rx}
J.~D. Lewin and P.~F. Smith, \emph{{Review of mathematics, numerical factors,
  and corrections for dark matter experiments based on elastic nuclear
  recoil}},
  \href{http://dx.doi.org/10.1016/S0927-6505(96)00047-3}{\emph{Astropart.
  Phys.} {\bf 6} (1996) 87--112}.

\bibitem{Altmannshofer:2014pba}
W.~Altmannshofer, S.~Gori, M.~Pospelov and I.~Yavin, \emph{{Neutrino Trident
  Production: A Powerful Probe of New Physics with Neutrino Beams}},
  \href{http://dx.doi.org/10.1103/PhysRevLett.113.091801}{\emph{Phys. Rev.
  Lett.} {\bf 113} (2014) 091801}, [\href{http://arxiv.org/abs/1406.2332}{{\tt
  1406.2332}}].

\bibitem{Ge:2017poy}
S.-F. Ge, M.~Lindner and W.~Rodejohann, \emph{{Atmospheric Trident Production
  for Probing New Physics}},
  \href{http://dx.doi.org/10.1016/j.physletb.2017.06.020}{\emph{Phys. Lett.}
  {\bf B772} (2017) 164--168}, [\href{http://arxiv.org/abs/1702.02617}{{\tt
  1702.02617}}].

\bibitem{Acciarri:2014isz}
{\scshape ArgoNeuT} collaboration, R.~Acciarri et~al., \emph{{Measurements of
  Inclusive Muon Neutrino and Antineutrino Charged Current Differential Cross
  Sections on Argon in the NuMI Antineutrino Beam}},
  \href{http://dx.doi.org/10.1103/PhysRevD.89.112003}{\emph{Phys. Rev.} {\bf
  D89} (2014) 112003}, [\href{http://arxiv.org/abs/1404.4809}{{\tt
  1404.4809}}].

\bibitem{Aliaga:2015wva}
{\scshape MINERvA} collaboration, T.~Le et~al., \emph{{Single Neutral Pion
  Production by Charged-Current $\bar{\nu}_\mu$ Interactions on Hydrocarbon at
  $\langle E_\nu \rangle = $3.6 GeV}},
  \href{http://dx.doi.org/10.1016/j.physletb.2015.07.039}{\emph{Phys. Lett.}
  {\bf B749} (2015) 130--136}, [\href{http://arxiv.org/abs/1503.02107}{{\tt
  1503.02107}}].

\bibitem{Acciarri:2015ncl}
{\scshape ArgoNeuT} collaboration, R.~Acciarri et~al., \emph{{Measurement of
  $\nu_{\mu}$ and $\bar{\nu}_{\mu}$ neutral current $\pi^{0} \rightarrow
  \gamma\gamma$ production in the ArgoNeuT detector}},
  \href{http://dx.doi.org/10.1103/PhysRevD.96.012006}{\emph{Phys. Rev.} {\bf
  D96} (2017) 012006}, [\href{http://arxiv.org/abs/1511.00941}{{\tt
  1511.00941}}].

\bibitem{Caratelli:2018nob}
D.~Caratelli, \emph{{Study of Electromagnetic Interactions in the MicroBooNE
  Liquid Argon Time Projection Chamber}}.
\newblock PhD thesis, Columbia U., 2018.
\newblock 10.2172/1420402.

\bibitem{duneprism}
{J. Cai et al.},
  ``{\url{https://indico.fnal.gov/event/16205/contribution/2/material/0/0.pdf}}.''

\end{thebibliography}\endgroup

\end{document}